\documentclass[9pt,twocolumn,twoside]{opticajnl}

\journal{opticajournal} 

\setboolean{shortarticle}{true}

\title{Real-time surrogate modeling of nonlinear pulse evolution in multimode fibers}
\author[ ]{Bora Çarpınlıoğlu}
\author[ ]{Bahadır Utku Kesgin}
\author[*]{Uğur Teğin}

\affil[ ]{Department of Electrical and Electronics Engineering, Koç University, Istanbul, Türkiye}

\affil[*]{utegin@ku.edu.tr}

\begin{abstract}
Modeling nonlinear pulse propagation in multimode fibers is challenging due to the large number of interacting modes and the resulting spatiotemporal complexity. Traditional optimization methods often become intractable, while learning-based approaches, such as recurrent neural networks, suffer from high computational cost and long inference times. We present a U-Net architecture as a fast, accurate surrogate for modeling nonlinear pulse propagation in multimode fibers. This approach overcomes the intractability of traditional methods while offering low computational cost. Trained on data generated by beam propagation method, our approach achieves an $\sim$88\% average structural similarity index with simulations. The model can generalize to untrained propagation distances, demonstrating convolutional architectures as efficient tools for simulating complex spatiotemporal dynamics in multimode fibers and offering potential for applications like mode decomposition.
\end{abstract}

\setboolean{displaycopyright}{false} 

\begin{document}

\maketitle

\smallskip

\noindent Optical fibers form the backbone of numerous modern technologies, with critical roles in applications ranging from advanced imaging techniques \cite{skarsoulis2024ptychographic} and novel computing paradigms \cite{oguz2024programming} to high-speed data transmission \cite{miller2012optical}. While single-mode fibers (SMFs) offer straightforward modeling and utilization due to their single propagation pathway, multimode fibers (MMFs) present a far richer landscape of spatiotemporal dynamics. This complexity, arising from the interplay of multiple guided spatial modes, makes MMFs highly attractive for innovative applications and exceptionally challenging for predictive modeling and experimental control \cite{wright2017multimode}. MMFs typically support over a hundred guided modes, a characteristic that significantly complicates their detailed analysis and can impede their straightforward deployment in practical optical systems.

Two main strategies have emerged to harness the potential of MMFs. The first embraces modal complexity, leveraging high-dimensional spatiotemporal interactions for applications such as optical computing, including photonic implementations of extreme learning machines \cite{kesgin2025photonic, iskandar2025optical, manuylovich2025optical}. The second focuses on understanding and controlling these dynamics through experimental or algorithmic techniques, including light control for imaging \cite{cao_controlling_2023}, suppression of nonlinear instabilities \cite{wisal_optimal_2024}, adaptive wavefront shaping \cite{tzang2018adaptive}, and single-mode output from mode-locked MMF lasers \cite{teugin2020single}.

The rise of deep learning has brought powerful data-driven frameworks to photonics, offering effective tools for modeling and controlling complex optical systems. The wide range of neural network architectures enables tailored solutions for specific photonic challenges. Comprehensive reviews have helped classify these models and guide experimental design \cite{freire_artificial_2023, genty_machine_2021}. In MMFs, deep learning has enabled notable advances, including feed-forward networks for managing spatiotemporal nonlinearities \cite{tegin_controlling_2020} and techniques for ultrafast pulse characterization \cite{xiong_deep_2020}.

Among these architectures, recurrent neural networks (RNNs) have garnered considerable attention for their efficacy in modeling nonlinear pulse propagation in optical fibers \cite{salmela_predicting_2021, tegin_reusability_2021}. This suitability stems from the natural conceptual alignment between the temporal, sequential evolution of optical pulses and the inherent memory dynamics characteristic of RNNs. The theoretical underpinnings for the aptitude of RNNs in modeling complex dynamical systems have also been robustly established \cite{jiang_model-free_2019}. However, despite their impressive modeling capabilities, RNNs are often hampered by substantial computational demands during training and inference phases. These resource-intensive requirements can significantly limit their practical applicability for tasks demanding real-time control or rapid optimization of spatiotemporal nonlinear interactions within MMFs, thereby motivating the search for more efficient alternatives.

\begin{figure}[ht]
\vspace{0.25cm}
\centering
\includegraphics[width=0.95\linewidth]{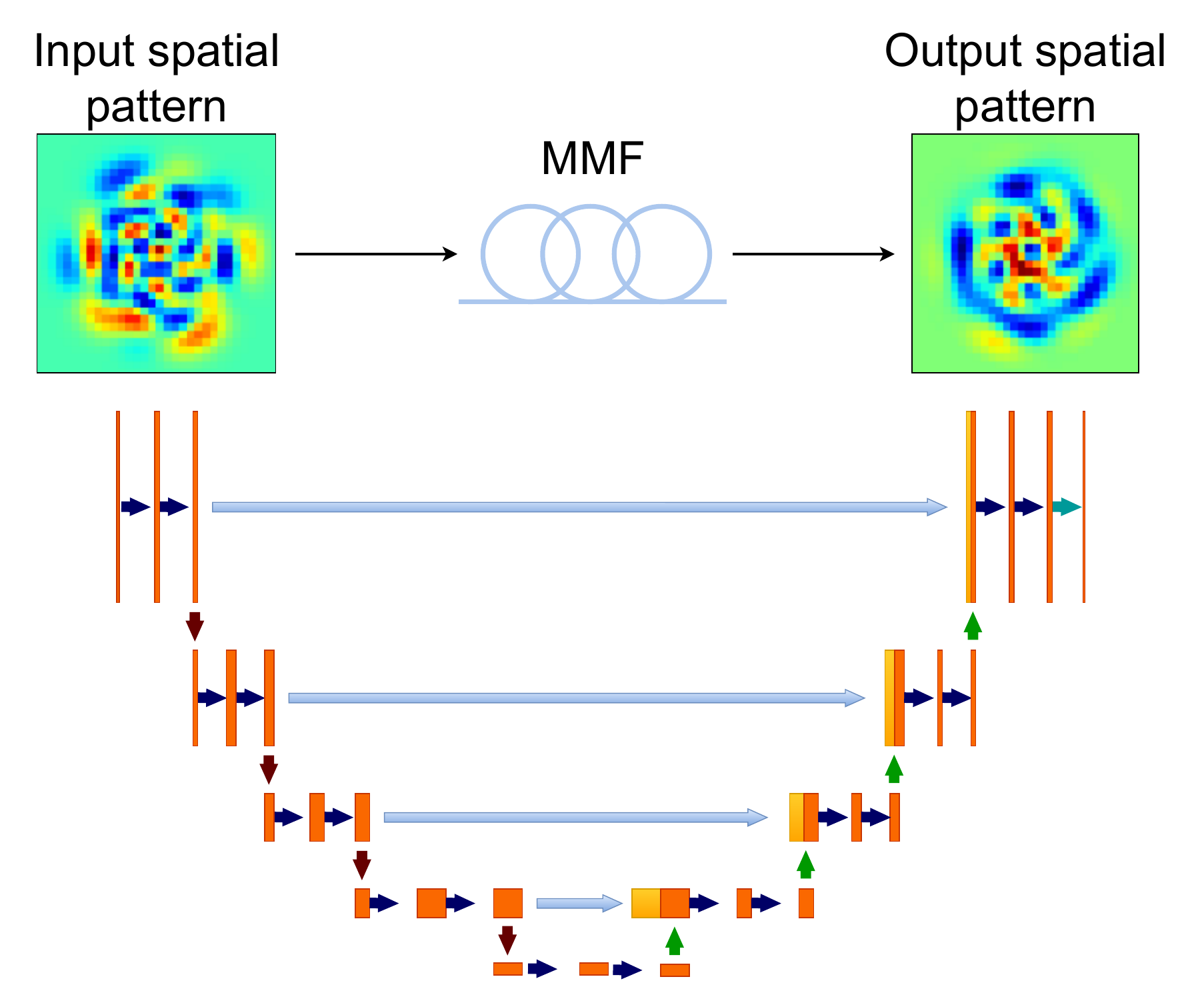}
\caption{The overall scheme of dataset generation and fitting a U-Net model to the generated dataset. A multimode fiber (MMF) is used to map input spatial patterns consisting of mode combinations to output spatial patterns.}
\label{fig:setup}
\end{figure}

In this work, we address the challenge of efficiently modeling nonlinear pulse propagation in MMFs by proposing a more practical alternative to computationally intensive RNNs. We focus on the U-Net architecture, a type of convolutional neural network, and demonstrate its capabilities by training it on a custom-generated dataset of input-output spatiotemporal field profiles from MMFs. Our results show strong agreement between the U-Net predictions and data from beam propagation method (BPM) simulations, as quantified by high structural similarity index metrics (SSIM). Furthermore, we investigate the model's generalization performance, assessing its ability to predict outcomes for longer propagation distances by cascading multiple trained U-Nets. Finally, by inverting the learned input-output relationship of our deep learning model, we demonstrate the U-Net's potential for predicting the necessary input field patterns to achieve specific target output patterns, opening a promising avenue for applications such as mode decomposition-based analyses and tailored pulse shaping.

\noindent
\textbf{Dataset generation }To generate a dataset exhibiting the rich spatiotemporal dynamics essential for training our deep learning model, we focused on the phenomenon of spatiotemporal instability (STI) in MMFs \cite{krupa_observation_2016, teugin2017spatiotemporal}. For a supervised learning approach, specific fiber parameters, detailed in Supplement 1, were selected and maintained consistently throughout the entire dataset generation procedure. We employed the beam propagation method (BPM) to perform simulations of light propagation in MMFs, thereby constructing the dataset for subsequent use with our U-Net architecture. The evolution of the complex field envelope, $A(x,y,z,t)$, within the multimode fiber is governed by the generalized nonlinear Schrödinger equation (GNLSE), written as:

\begin{multline}
  \frac{\partial A}{\partial z} = \frac{i}{2k_0}\left(\frac{\partial^2A}{\partial x^2} + \frac{\partial^2A}{\partial y^2}\right) - i\frac{\beta_2}{2}\frac{\partial^2 A}{\partial t^2} \\ + \frac{\beta_3}{6}\frac{\partial^3 A}{\partial t^3} - \frac{ik_0\Delta(x^2+y^2)A}{R^2} + i\gamma {|A|}^2A
   \label{eqs:gnlse} 
\end{multline}
where $k_0$ is the wavenumber, $\beta_2$ and $\beta_3$ are the second- and third-order dispersion coefficients, respectively, $\Delta$ relates to the refractive index difference for a graded-index profile of radius $R$, and $\gamma$ is the nonlinear coefficient. 

Numerical solution of Eq.~(\ref{eqs:gnlse}) was performed using the symmetrized split-step Fourier method (SSFM). For each of the $N$ dataset elements, the input field $A_{in}(x,y,t)$ (at $z=0$) was constructed as an equal linear combination of five spatial modes randomly selected from a total of 120 modes supported by the MMF. This input field was then propagated for a distance of 2.5 cm with an initial peak power of 5 MW. These parameters were chosen to induce significant nonlinear spatiotemporal evolution over a relatively short propagation distance, providing diverse and complex output fields. After each propagation, the initial input field and the resulting output field at $z=2.5 \text{ cm}$ were recorded. From these full spatiotemporal fields, corresponding pairs of input and output two-dimensional spatial beam profiles were extracted to form the dataset. This simulation campaign resulted in 4250 such input-output profile pairs, with the entire dataset generation process requiring approximately 42 hours of computation time.

\noindent
\textbf{Deep learning architecture } Choosing an appropriate neural network architecture for analyzing the large datasets generated in ultrafast photonics can be challenging. While general guidelines exist—feed-forward networks are often suitable for single-pass regression or classification tasks, and various convolutional architectures are preferred for data with inherent spatial or spectral structure \cite{genty_machine_2021}—systems with strong temporal dependencies frequently employ recurrent neural networks (RNNs) \cite{salmela_predicting_2021, tegin_reusability_2021}. However, as our goal is to efficiently map input spatial field profiles to output profiles from the multimode fiber (MMF), analogous to an image-to-image translation task, we selected the U-Net architecture \cite{ronneberger2015u}.

\begin{figure}[b!]
\centering\includegraphics[width=\linewidth]{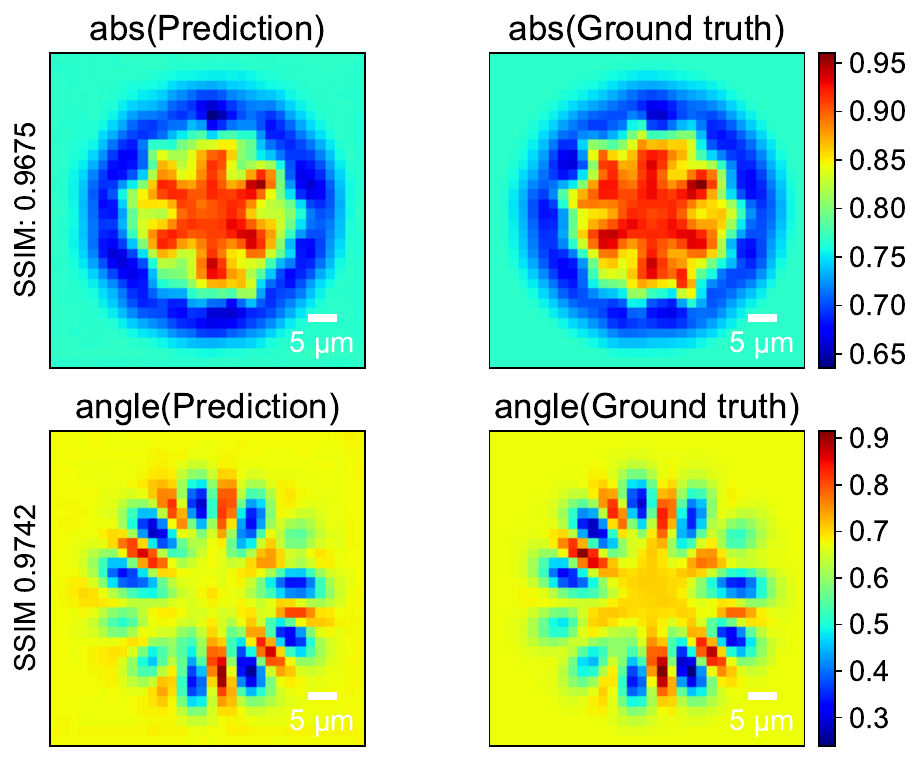}
\caption{U-Net training results for 2.5 cm propagation in a multimode fiber. The amplitude (top) and phase (bottom) parts of predicted and ground truth complex fields are compared together with corresponding structural similarity index values.}
\label{fig:dnn3}
\end{figure}

The U-Net, a specialized fully convolutional neural network, is particularly well-suited for this purpose due to its proven efficacy in image segmentation and transformation tasks. Its encoder-decoder structure with skip connections allows it to capture both contextual information and fine-grained spatial details, crucial for resolving the complex patterns generated during nonlinear propagation in MMFs. This architecture offers robust pattern-finding capabilities while significantly mitigating the long training and inference times often associated with RNNs, a key consideration for our work. A further practical advantage is its adaptability to complex-valued data; standard U-Net implementations, typically designed for real-valued inputs, can readily process complex electromagnetic fields by treating their real and imaginary components (or alternatively, amplitude and phase) as two separate channels in the input and output tensors. Our specific U-Net implementation and its integration into the simulation and analysis workflow are depicted in Fig.~\ref{fig:setup}.

\begin{figure}[ht!]
\centering\includegraphics[width=0.93\linewidth]{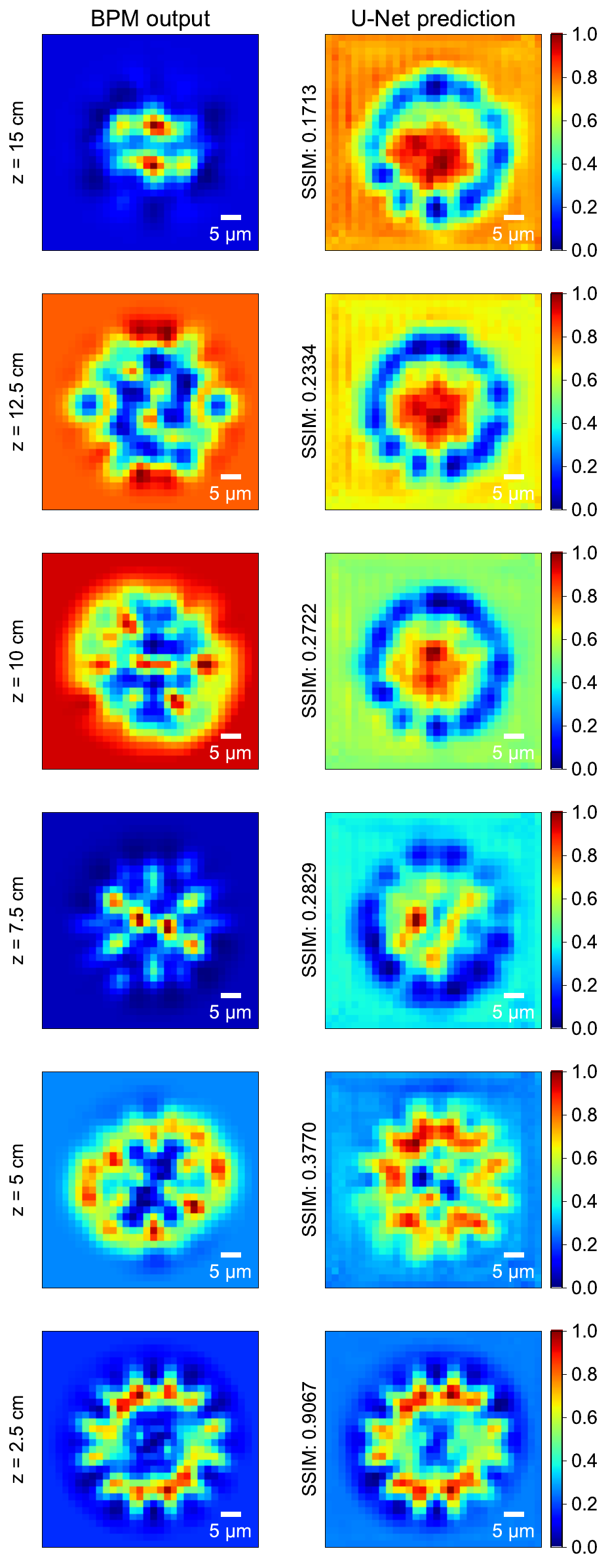} 
\caption{Normalized intensity profiles obtained from BPM simulations (left) and by cascading the trained U-Nets (right) six times illustrated at integer multiples of the initial propagation distance, 2.5 cm. SSIM values are given to the left of U-Net predictions.}
\label{fig:cascade}
\end{figure}

For training and evaluating the U-Net model, the dataset was partitioned into training and testing sets using an 80/20 split, respectively. To optimize the network's performance for our specific task, we leveraged the automatic hyperparameter search capabilities of the Ray Tune library \cite{liaw2018tune} for each training scenario explored in this study.

\noindent
\textbf{Results} The U-Net architecture was trained using the generated dataset, achieving an average mean square error (MSE) of $0.1\%$ on the test set for the primary task of forward propagation prediction. While MSE provides a basic measure of error, the structural similarity index measure (SSIM) offers more perceptually relevant insights into the model's performance, especially for image-like data. SSIM is a widely adopted metric in deep learning for image-to-image tasks and has recently seen increased application in optics \cite{anisimov2023similarity}. It quantifies the similarity between predicted and ground-truth spatial profiles. For our analysis, we computed SSIM values separately for the absolute (amplitude) and phase components of the complex spatial beam profiles in the test set. As shown by representative examples in Fig.~\ref{fig:dnn3}, which visualizes learning results for a sample from the test set alongside their SSIM values, the model demonstrates high fidelity. The average SSIM values obtained were $0.8823$ for the amplitude and $0.8909$ for the phase. These results indicate that our trained U-Net can serve as an effective, high-fidelity surrogate for computationally intensive BPM simulations under fixed fiber and input power parameters.

A critical aspect of any predictive model is its generalization capability. We investigated our U-Net's ability to extrapolate and predict propagation outcomes for distances exceeding 2.5 cm used during training. This was achieved by cascading the trained U-Net—using the output of one U-Net as the input for the next identical, independently operating U-Net—to simulate longer propagation paths, and comparing these predictions against new ground-truth BPM simulations. This cascading approach is viable because our U-Net processes and preserves both channels (representing amplitude and phase) of the complex field. Figure~\ref{fig:cascade} illustrates the evolution of representative spatial profiles (e.g., intensity) at six successively increasing $z$ positions, along with corresponding SSIM values for the intensity at each step. As anticipated, the SSIM values gradually decrease with increasing propagation distance, signifying a reduction in prediction accuracy for further extrapolations. Nevertheless, this cascaded model offers a substantial speed advantage over BPM simulations, making it a valuable tool for applications requiring real-time or near-real-time feedback over extended propagation distances, even if with a trade-off in absolute accuracy.

\begin{figure}[t!]
\centering\includegraphics[width=\linewidth]{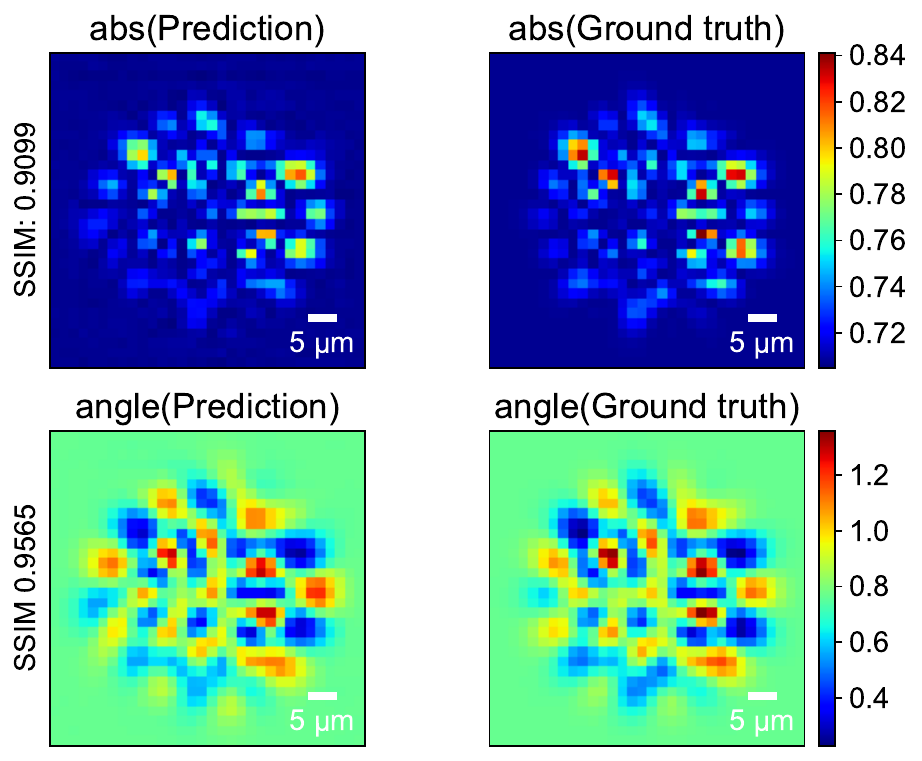}
\caption{U-Net training results for the inverse problem where the roles of the input and output are reversed. The amplitude (top) and phase (bottom) parts of predicted and ground truth complex fields are compared together with corresponding structural similarity index values.}
\label{fig:inverse}
\end{figure}

Leveraging the U-Net's capability as a versatile function approximator, we also explored its application to the inverse problem: predicting the necessary input spatial field profile at $z=0$ that would result in a desired target output profile at $z=2.5 \text{ cm}$. This required training a separate U-Net model with the roles of input and output in our dataset reversed. Such predictive capability for input fields could be highly beneficial for tasks like mode decomposition or targeted excitation of specific nonlinear phenomena \cite{rothe2020deep}. For this inverse prediction task, the retrained U-Net achieved an average MSE of $0.17\%$ on its test set. The average SSIM values were $0.6998$ for the amplitude and $0.7828$ for the phase. While these SSIM values are lower than those for the forward propagation model, indicating the increased complexity of the inverse problem, the qualitative results (exemplified in Fig.~\ref{fig:inverse}) demonstrate considerable promise. These findings suggest that a U-Net-based inverse model can be a valuable tool for exploring and potentially controlling input conditions to achieve specific outcomes in nonlinear MMF systems, for instance, by helping to identify the initial modal compositions that lead to desired nonlinear effects.

Compared to prior deep learning approaches for spatiotemporal modeling in MMFs, such as recurrent \cite{tegin_reusability_2021, salmela_predicting_2021} and feed-forward neural networks \cite{genty_spie}, our U-Net-based method offers substantially reduced inference times (detailed in Supplement 1). This positions it as a promising candidate for applications demanding real-time or near-real-time performance. While the current study is computational, and experimental validation faces the significant challenge of precise modal excitation in MMFs, we anticipate our findings will be valuable for advancing both numerical explorations and, ultimately, experimental control of spatiotemporal nonlinearities.

Several exciting avenues emerge from this work. Architectural enhancements, incorporating spatiotemporal convolutions inspired by video analysis techniques, could further improve modeling intricate dynamics \cite{tran_closer_2018}. Our surrogate model’s efficiency suggests its applicability in computationally intensive tasks beyond MMFs, such as inverse design of integrated photonic components where fast and accurate simulation is critical \cite{molesky2018inverse}. Our dataset and inverse modeling capability pave the way for future studies on specific spatial modes contributing to nonlinear phenomena. These results highlight the potential of tailored deep learning architectures for accelerating discovery and enabling new functionalities in complex nonlinear photonic systems.

In conclusion, we have demonstrated a U-Net architecture as an efficient and accurate surrogate for modeling nonlinear pulse propagation in multimode fibers, achieving >0.88 average SSIM with BPM simulations for predicting spatial field profiles. The model generalizes to longer propagation distances and successfully performs inverse prediction of input fields from outputs, tasks critical for understanding and control. Offering substantially reduced inference times compared to established deep learning methods, this U-Net approach is highly promising for real-time applications, advanced numerical studies, and future inverse design problems in photonics. Our findings highlight the significant potential of tailored deep learning to address complex challenges in nonlinear optical systems.

\begin{backmatter}

\bmsection{Disclosures} The authors declare no conflicts of interest.

\bmsection{Data Availability Statement} 
\noindent
Data underlying the results presented in this paper are not publicly available at this time but may be obtained from the authors upon reasonable request.

\bmsection{Supplemental document} See Supplement 1 for supporting content.
\end{backmatter}

\bibliography{modal_excitation}

\bibliographyfullrefs{modal_excitation}

\end{document}